\documentclass[a4paper,fleqn,longmktitle]{cas-sc}  
\usepackage[authoryear]{natbib}

\def\tsc#1{\csdef{#1}{\textsc{\lowercase{#1}}\xspace}}
\tsc{WGM}
\tsc{QE}
\tsc{EP}
\tsc{PMS}
\tsc{BEC}
\tsc{DE}

\begin{document}

\let\WriteBookmarks\relax
\def\floatpagepagefraction{1}
\def\textpagefraction{.001}
\shorttitle{KAW in the Magnetosphere}
\shortauthors{Moya et~al.}

\title[mode = title]{The effect of oxygen ions on the stability and polarization of Kinetic Alfv\'en Waves in the Magnetosphere}                      

\author[1,2]{Pablo S. Moya}[orcid=0000-0002-9161-0888]
\ead{pablo.moya@uchile.cl}
\cormark[1]

\author[1]{Iv\'an Gallo-M\'endez}[orcid=0000-0003-4988-4348]

\author[1]{Bea Zenteno-Quinteros}[orcid=0000-0003-2430-6058]

\address[1]{Departamento de F\'isica, Facultad de Ciencias, Universidad de Chile, Santiago, Chile}
\address[2]{Centre for mathematical Plasma Astrophysics, KU Leuven, Leuven, Belgium}

\cortext[cor1]{Corresponding author}

\begin{abstract}
One of the most striking properties of Kinetic Alfv\'en Waves (KAW) 
is that, unlike the also Alfv\'enic Electromagnetic Ion Cyclotron (EMIC) waves, these waves are right-hand polarized in the plasma frame. In particular, this signature property is key for the identification of KAW from in situ measurements of plasma waves. From the theoretical point of view, both the dispersion relation and the polarization of KAW has been mostly studied in proton-electron plasmas. However, most astrophysical and space plasmas are multi-species, and therefore in these systems the dispersion properties of the KAW may not depend only on the macroscopic parameters of proton and electron distributions, but also on the parameters of heavier ions. Here, using Vlasov linear theory we study the dispersion properties of Alfv\'enic modes in multi-species plasmas composed by electrons, protons, and O$^+$ ions, with macroscopic plasma parameters relevant to the inner magnetosphere. In consistency with recent observations, our numerical results show that the presence of O$^+$ ions allows the existence of KAW in a wider wave-number range and at smaller wave-normal angles compared to the electron-proton case, but at the same time isotropic O$^+$ ions tend to reduce (or even inhibiting) the growth rates of unstable KAW triggered by anisotropic protons. These results suggest that magnetospheric ions may play an important role on the energy transfer from large macroscopic scales to sub-ionic and electronic scales, especially during intense geomagnetic storms in which O$^+$ ions can dominate the plasma composition in the inner magnetosphere. 
\end{abstract}



\begin{keywords}
Multi-species space plasmas \sep Magnetospheric O$^+$ ions \sep Plasma waves \sep Kinetic Alfv\'en Waves 
\end{keywords}

\maketitle

\section{Introduction}
\label{ref:intro}

Kinetic Alfv\'en Waves (KAW) are Alfv\'enic waves that appear when, at large wave-normal angles, kinetic effects of electrons become relevant~\citep{hasegawa1976,gary1986,hollweg1999}, and are of particular interest on the study of turbulence in space and astrophysical environments~\citep{Boldyrev2013,Narita2020}. As in general KAW wave-particle interactions are resonant with electrons and non-resonant with ions, KAW have been proposed as a possible mechanism that allows electromagnetic energy transference from the scales in which wave-particle interactions with ions and protons dominate, towards the smaller and faster electron scales~\citep{nandal2016}. There has been a wide interest in study the role of these waves in space plasmas such as the solar wind~\citep{salem2012,Pucci2016,marsch2018}, and the magnetospheric environment. In the case of the Earth's magnetosphere, KAW has been observed in the magnetosheath~\citep{Chasapis2018,Macek2018,Dwivedi2019}, the plasma sheet and plasma sheet boundary layer~\citep{wygant2002,chaston2012} and the inner magnetosphere~\citep{chaston2014,moya2015}. Furthermore, KAW has been found to play an important role in several kinetic processes such as ion and electron heating~\citep{wygant2002,chaston2014}, energization of the ionosophere~\citep{Keiling2019,Cheng2020}, ionospheric outflow~\citep{chaston2006}, and reconnection~\citep{chaston2005,Boldyrev2019}, among others. 

One of the most remarkable characteristics of KAW, unlike the also Alfv\'enic left-hand polarized Electromagnetic Ion-Cyclotron (EMIC) waves, is that they are right-handed in the plasma frame as discovered by~\citet{gary1986}. Also according to~\citet{gary1986}, the change in polarization is related with the development of large parallel electric field, such that the electric field becomes primarily electrostatic. To do so, KAW requires the ion gyroradius $\rho_i$ to be of the order of the wavelength ($k_{\perp} \rho_i \sim 1$), where $k_\perp$ is the wave-number in the direction perpendicular to the background magnetic field, such that the mode becomes linearly compressive (EMIC waves are non-compressive modes) and develops a large parallel electric field~\citep{hollweg1999}. Therefore, the wave-normal angle in which the Alfv\'enic waves shift from EMIC to KAW is highly dependent on plasma beta~\citep{gary1993,Gary2004}. From the theoretical point of view, the dispersion properties of KAW have in general been studied in proton-electron plasmas~\citep[see e.g.][and references therein]{gary1986,hollweg1999,Lopez2017}, however, most astrophysical and space plasmas are multi-species. Then, dispersion properties such as polarization should also depend on the ionic species different than the protons. As the presence of other ions must alter the susceptibility of the media, especially at frequencies and wave-numbers similar to the gyrofrequency, gyroradius or inertial length of the particles, it is expected the dispersion relation and fluctuating fields to be also altered. Therefore, the polarization of the waves should also be influenced. 

As each ion species has its own gyrofrequency, in a multi-ion plasma wave-particle interactions can be resonant (or non-resonant) with protons but also with other ion species. Subsequently, the local stability of a given wave-mode will depend on macroscopic properties (e.g. plasma beta, temperature anisotropy or bulk velocity) of protons, electrons and the rest of ion species~\citep[see e.g.][and references therein]{moya2014,Yoon2017}. The free energy for micro-instabilities can be stored by one or several species; e.g, a temperature anistropy driven instability can be triggered by only one anisotropic species even though all other species are isotropic. Under this context, it has recently been shown that even in absence of temperature anisotropic protons, drifting ions or velocity shear can generate unstable Kinetic Alfv\'en Wave~\citep{Lakhina2008,Barik2019a,Barik2019b}, and that supra-thermal electrons can also play an important role on the properties of KAW modes~\citep{Barik2019c,Barik2020}. Further, as the effectiveness and growth rates of resonant instabilities depend strongly on the portion of the plasma particles undergoing resonant wave-particle interactions~\citep{stix1992}, then the relative abundance of each species should also play a role on the determination of the maximum growth of the instability, even inhibiting the instability if the abundance of the unstable species is not large enough.

These issues may be especially relevant in the inner magnetosphere as there is a strong dependence between the relative abundance of protons, and O$^+$ ions and geomagnetic activity. It has been shown that due to ionospheric outflow, during geomagnetic storms O$^+$ ions can even dominate the composition of the plasma in the ring current region~\citep{hamilton1988,daglis1999,Yue2018}, and that the abundance of each ion species is highly dependent on the intensity of the storm~\citep{jahn2017,Denton2017}. Thus, it is expected the changes on the plasma composition due to geomagnetic activity to enhance or suppress the propagation or stability of KAW, and therefore to play a role on the kinetic processes mediated by these waves, as it has been shown considering plasma parameters relevant to the vicinity of comets~\citep{Samuel2011,Venugopal2014} or the cusp region in the magnetosphere~\citep{Tamrakar2017}.

In this article, motivated by recent works that have reported observations of KAW modes in the inner magnetosphere propagating at about $60^\circ$ with respect to the magnetic field~\citep{chaston2014,moya2015}, considering plasma parameters relevant for the ring current region, we present a study on the effect of O$^+$ ions on the dispersion relation of Alfv\'enic waves propagating at a oblique angle with respect to the background magnetic field, analyzing the changes that these heavy ions introduce on the dispersion relation, stability and polarization (all as function of wave-number) of Kinetic Alfv\'en Waves when compared with the electron-proton case. To the best of our knowledge, unlike previous works in which fluid models~\citep{Yang2005} or approximate versions of the kinetic dispersion relation~\citep{Samuel2011,Venugopal2014,Tamrakar2017} were studied, here for the first time we consider the full dispersion tensor with no approximations and choose on wave-normal angles $\theta<70^\circ$ to focus on the transition between left-handed to right-handed solutions and the role of the O+ ions on the properties of the waves. In the next section we briefly present the basic equations to obtain the Vlasov dispersion relation and polarization of KAW in a warm multi-species plasma in which each species follows a bi-Maxwellian velocity distribution function. Then, in Section~\ref{ref:results}, considering a plasma composed by electrons, protons, and O$^+$ ions with parameters relevant to the inner magnetosphere, we compute the complex dispersion relation and polarization spectra for Alfv\'enic waves at different wave-normal angle, comparing the results obtained in an electron-proton plasma, and the results when O$^+$ ions are considered. Finally, in section~\ref{ref:conclusions} we summarize our results and outline the main conclusions.

\section{Linear theory: Vlasov-Maxwell waves in multi-species plasmas}
\label{ref:linear}

For our analysis we consider a multi-species warm plasma, with a uniform background magnetic field $\mathbf{B}_0 = B_0 \hat z$, in which each species $s$ follows a drifting bi-Maxwellian velocity distribution function: 

\begin{equation}
\label{eq:vdf}
    f_s(v_\bot, v_\parallel)=\frac{n_{0 s}}{\pi^{3/2}\alpha^2_{\bot s}\alpha_{\parallel s}}\exp\left(-\frac{v^2_\perp}{\alpha^2_{\bot s}}-\frac{\left(v_\parallel-U_{\parallel s}\right)^2}{\alpha^2_{\parallel s}}\right)\,,
\end{equation}
where $\alpha^2_{\bot s} = 2k_BT_\bot/m_s$ and $\alpha^2_{\parallel s} = 2k_BT_\parallel/m_s$ are the squares of the thermal speeds of the species $s$, and $T_{\bot s}$ and $T_{\parallel s}$ are the perpendicular and parallel temperatures with respect to $\mathbf{B}_0$, respectively. Also, in Eq. \eqref{eq:vdf}, $n_{0s}$, $m_s$, and $U_{\parallel s}$ denote the number density, mass, and drift velocity along the magnetic field of the $s$th species, and $k_B$ is the Boltzmann constant. In such plasma, the Vlasov-Maxwell linear kinetic dispersion relation for electromagnetic waves is determined by
\begin{equation}
\label{eq:DR}
    D(\omega,\,\mathbf{k})\,\delta\mathbf{E_k}(\omega, \mathbf{k}) = 0\,
\end{equation}
or $|D(\omega,\,\mathbf{k})| = 0$, where $\omega = \omega(\mathbf{k}) = \omega_r(\mathbf{k}) + i\,\gamma(\mathbf{k})$ are the complex solutions of the dispersion relation (eigen-frequencies) as a function of the wave-vector $\mathbf{k}$. Here, for simplicity, and without loss of generality, we have restricted the wave-vector to the $x$--$z$ plane, so that $\mathbf{k} = k_\bot \hat x + k_\parallel \hat z$, and the wave-normal angle $\theta$ is given by $\tan(\theta) = k_\bot /k_\parallel$. In Eq.~\eqref{eq:DR} $\delta\mathbf{E_k}$ corresponds to the fluctuating electric field (eigen-modes) of the plasma, and the dispersion tensor $D$ is given by~\citep{stix1992,vinas2000,Lopez2017,Yoon2017}

  \begin{equation}
  \label{eq:lambda}
    D(\omega, \mathbf{k}) = 
  \begin{pmatrix}
 D_{xx}& D_{xy}  & D_{xz} \\ 
-D_{xy} & D_{yy} & D_{yz} \\ 
 D_{xz}&  -D_{yz} & D_{zz}
\end{pmatrix} \,,
  \end{equation}
with
\begin{equation}
\label{eq:dxx}
D_{xx} = 1 - \frac{k_{\parallel}^2c^2}{\omega^2} + 2\sum_s \sum_{n=-\infty}^{\infty} \frac{\omega_{ps}^2}{\omega^2} \Lambda_n(\lambda_s) \left(\frac{n^2 \Omega_s^2}{k_{\bot}^2\alpha_{\bot s}^2}\right) [(\mu_s - 1) +\mu_sZ(\xi_{ns})\bar{\xi}_{ns}]\,,    
\end{equation}

\begin{equation}
D_{xy} = i\sum_s \sum_{n=-\infty}^{\infty} \frac{\omega_{ps}^2}{\omega^2}\mu_s n \Lambda'(\lambda_s) Z(\xi_{ns}) \bar{\xi}_{ns}\,,    
\end{equation}

\begin{equation}
D_{xz} = \frac{k_{\parallel}k_{\bot}c^2}{\omega^2} + 2\sum_s\sum_{n=-\infty}^{\infty} \frac{\omega_{ps}^2}{\omega^2} \Lambda_n(\lambda_s)\left(\frac{n\Omega_s}{k_{\bot}\alpha_{\parallel s}}\right)\left[\left(\mu_s^{-1} - 1\right) \left(\frac{n\Omega_s}{k_{\parallel}\alpha_{\parallel s}}\right) + y_{ns} Z(\xi_{ns}) \bar{\xi}_{ns}\right]\,,    
\end{equation}

\begin{equation}
D_{yy} = 1 - \frac{k^2c^2}{\omega^2} + 2\sum_s\sum_{n=-\infty}^{\infty} \frac{\omega_{ps}^2}{\omega^2} \left(\frac{\Omega_s^2}{k_{\bot}^2\alpha_{\bot s}^2}\right) (n^2\Lambda_n(\lambda_s)-2\lambda_s^2\Lambda'_n(\lambda_s)) \left[(\mu_s - 1) + \mu_s Z(\xi_{ns}) \bar{\xi}_{ns}\right]\,,   
\end{equation}

\begin{equation}
D_{yz} = -2i\sum_s\sum_{n=-\infty}^{\infty} \frac{\omega_{ps}^2}{\omega^2} \Lambda'_n(\lambda_s)\left(\frac{\lambda_s\Omega_s}{k_{\bot}\alpha_{\parallel s}}\right)y_{ns} Z(\xi_{ns}) \bar{\xi}_{ns}\,,
\end{equation}

\begin{equation}
 D_{zz} = 1 - \frac{k_{\bot}^2c^2}{\omega^2} + 2\sum_s\sum_{n=-\infty}^{\infty} \frac{\omega_{ps}^2}{\omega^2} \Lambda_n(\lambda_s)\left[\left(\frac{\omega}{k_{\parallel} \alpha_{\parallel s}}\right)^2 + \left(1 - \mu_s^{-1}\right)\left(\frac{n\Omega_s}{k_{\parallel}\alpha_{\parallel s}}\right)^2 + y_{ns}^2 Z(\xi_{ns}) \bar{\xi}_{ns}\right]\,,   
\end{equation}
where $\omega_{ps} = (4\pi \,n_{0s}\, q^2/m_s)^{1/2}$, $\,\Omega_s = q_s B_0/m_s c$, and $q_s$ are the plasma frequency, gyrofrequency, and charge of the species $s$, $c$ is the speed of light; and we have defined the following parameters for each species $s$:
\begin{equation}
\mu_s = \frac{T_{\bot s}}{T_{\parallel s}}\,, \quad
\xi_{ns} = \frac{\omega - n\Omega_s - k_{\parallel}U_{\parallel s} }{k_{\parallel}\alpha_{\parallel s}}\,, \quad
\bar{\xi}_{ns} = \frac{\omega - n\Omega_s(1 - \mu_s^{-1}) - k_{\parallel}U_{\parallel s}}{k_{\parallel}\alpha_{\parallel s}}\,,\quad
y_{ns} = \frac{\omega - n\Omega_s }{k_{\parallel}\alpha_{\parallel s}}\,,
\end{equation}

\begin{equation}
\label{eq:bessel}
\lambda_s = \frac{1}{2}\frac{k_{\bot}^2 \alpha_{\bot s}^2}{\Omega_s^2}\,,\quad
\rm{and} \quad \Lambda_n(\lambda_s) = e^{-\lambda_s}I_n(\lambda_s)\,.
\end{equation}
Finally, in Eqs.~\eqref{eq:dxx}-\eqref{eq:bessel} $I_n$ corresponds to the modified Bessel function of order $n$, and $Z(x)$ is the so called Plasma Dispersion Function~\citep{friedandconte1961}. 

Besides the complex solutions $\omega(\mathbf{k})$, through the dispersion relation~\eqref{eq:DR} and Maxwell's equations it is also possible to obtain spectral information about each eigen-mode $\delta\mathbf{E_k}(\omega, \mathbf{k})$. Among others, the polarization spectrum, defined as~\citep{gary1986}

\begin{equation}
\label{eq:pol}
P(k) = i \frac{\delta E_{kx}}{\delta E_{ky}}\,,    
\end{equation}
can provide key information for the identification of KAW (right-hand polarized with $\rm{Re}\,[P]$ $>$ 0), and distinguish these modes from EMIC waves (left-handed and with $\rm{Re}\,[P]$ $<$ 0). 

Due to the explicit dependence of the elements of the dispersion tensor on the properties of each species composing the plasma, the solutions of the dispersion relation, Eq~.\eqref{eq:DR}, and the dispersion spectral properties of each wave-mode will vary depending on the composition of the plasma, the plasma parameters of each species, and the wave-normal angles. This dependence can be expressed in terms of a reduced number of dimensionless quantities such as the abundance $\eta_s = n_{0s}/n_0$ of the species $s$ with respect to the total density $n_0$, the parallel plasma beta $\beta_{\parallel s} = 8\pi n_s k_B T_{\parallel s}/B^2_0$, and the temperature anisotropy $\mu_s$ of each species. Also, global information about the particular environment of interest can be represented by the ratio between the Alfv\'en speed, $v_A = B_0/\sqrt{4\pi n_0 m_p}$, and the speed of light $C_A = v_A/c$ (where $m_p$ is the proton mass). All these parameters provide a way to characterize the linear response of the media to the propagation of electromagnetic waves at different wave-normal angle in different plasma environments.

\section{Numerical results}
\label{ref:results}

To numerically analyze the case of the inner magnetosphere we consider parameters relevant to the ring current region. In this region the plasma is composed mainly by electrons, H$^+$ ions, O$^+$ ions, and He$^+$ ions, with abundances that are strongly dependent on the level of geomagnetic activity. From all heavy ions species that can be found in the magnetosphere, it has been shown that depending on the strength of a given geomagnetic storm, O$^+$ ions can dominate the composition of the plasma, with $\eta_{\rm{O^+}}$ varying between 0.2 and 0.8~\citep{hamilton1988,daglis1999}, and that the abundance tend to increase with increasing geomagnetic activity. In contrast, the abundance of He$^+$ ions usually does not exceed 5\% during geomagnetically active times in the ring current region~\citep{jahn2017}, and therefore the helium ions are always the least abundant species. With this in mind, as a first approach to the study on the effect of heavy ions on the dispersion relation, stability and polarization of KAW in the inner magnetosphere, we consider a plasma composed by electrons, protons and O$^+$. We consider a quasi-neutral plasma in which $e(n_0-n_e) = 0$, where $e$ is the elementary charge, and $n_0$, $n_e$ are the total ion and electron number density, respectively. Then, $n_{\rm{H^+}} + n_{\rm{O^+}} = n_0 = n_e$, and solve the dispersion relation for different values of the O$^+$ abundance in the range $0 \leq \eta_{\rm{O^+}} \leq 0.40$, so that for each case $\eta_{\rm{H^+}} + \eta_{\rm{O^+}} = 1$. 

For the selection of the temperature anisotropy of each species, for simplicity, we consider a non-drifting ($U_s =0$) plasma, composed by isotropic electrons and O$^+$  ions $(\mu_e= \mu_{\rm{O^+}} = 1)$, and anisotropic protons that provide the free energy to destabilize the Alfv\'enic modes, with $0.8 \leq \mu_{\rm{H^+}} \leq 2.5$. Finally, for the parallel beta parameter we consider a temperature typical for the ring current $T\sim$~10 keV~\citep{kamide}, and total electron number density $n_e\sim10$~cm$^{-3}$ that has been measured in the region of interest, outside of the plasmasphere, at the same time that KAW modes were observed~\citep{moya2015}. With these density and temperature values, considering a magnetic field strength $B_0\sim$ 150 nT, then the parallel beta of each species will be $\beta_{\parallel s}\sim$~1. Therefore, for our study we consider $\bar\beta_{\parallel\, \rm{H^+}} = \bar\beta_{\parallel\, \rm{O^+}} = 1$, where $\bar\beta_s$ corresponds to the plasma $\beta_s$ of each species but considering the total density ($\bar\beta_s = 8\pi n_0 k_B T_s/B^2_0$), so that $\bar\beta_s$ is a measure of the temperature of each species as defined by~\citet{gary1993}. Finally, with these density and magnetic field values the ratio between the Alfv\'en speed and the speed of light is set to $C_A = 3.5\,\times\,10^{-3}$. It is important to mention that, as the important quantity controlling the shift between EMIC waves and KAW is plasma beta~\citep{gary1993,Gary2004}, due to the definition of beta, other combinations between density $n$, temperature $T$ and magnetic field $B$ can lead to the same value for the beta parameter, and therefore the same numerical results. 

To obtain solutions of the Vlasov linear dispersion, we use our own developed kinetic dispersion solver written in Python, that allows us to find the solutions of the dispersion relation [Eq.~\eqref{eq:DR}] at arbitrary wave-normal angle, and the dispersion properties of each eigen-mode such as the polarization [Eq.\eqref{eq:pol}] in multi-species plasma composed by bi-Maxwellian distributions. Using these numerical tools, in the two following sections we address and quantify the effect that O$^+$ ions introduce on the dispersion relation, stability and polarization of KAW. First, we solve the dispersion relation considering a simple electron-proton plasma, looking for the condition under which the Alfv\'enic modes correspond to unstable right-hand polarized KAW. Even though the properties of KAW modes are already well known, through this first analysis we perform a quick validation of our dispersion code, and corroborate previous theoretical~\citep{gary1986,hollweg1999,Gary2004} or observational~\citep{chaston2014,moya2015} results. More important, for the sake of the self-consistency and readability, the results presented in the next subsection provide plots to compare the electron-proton case with the multi-species case, when we introduce O$^+$ ions.

\subsection{Stability of KAW in an electron-proton plasma}

\begin{figure}
	\centering
		\includegraphics[width=.75\linewidth]{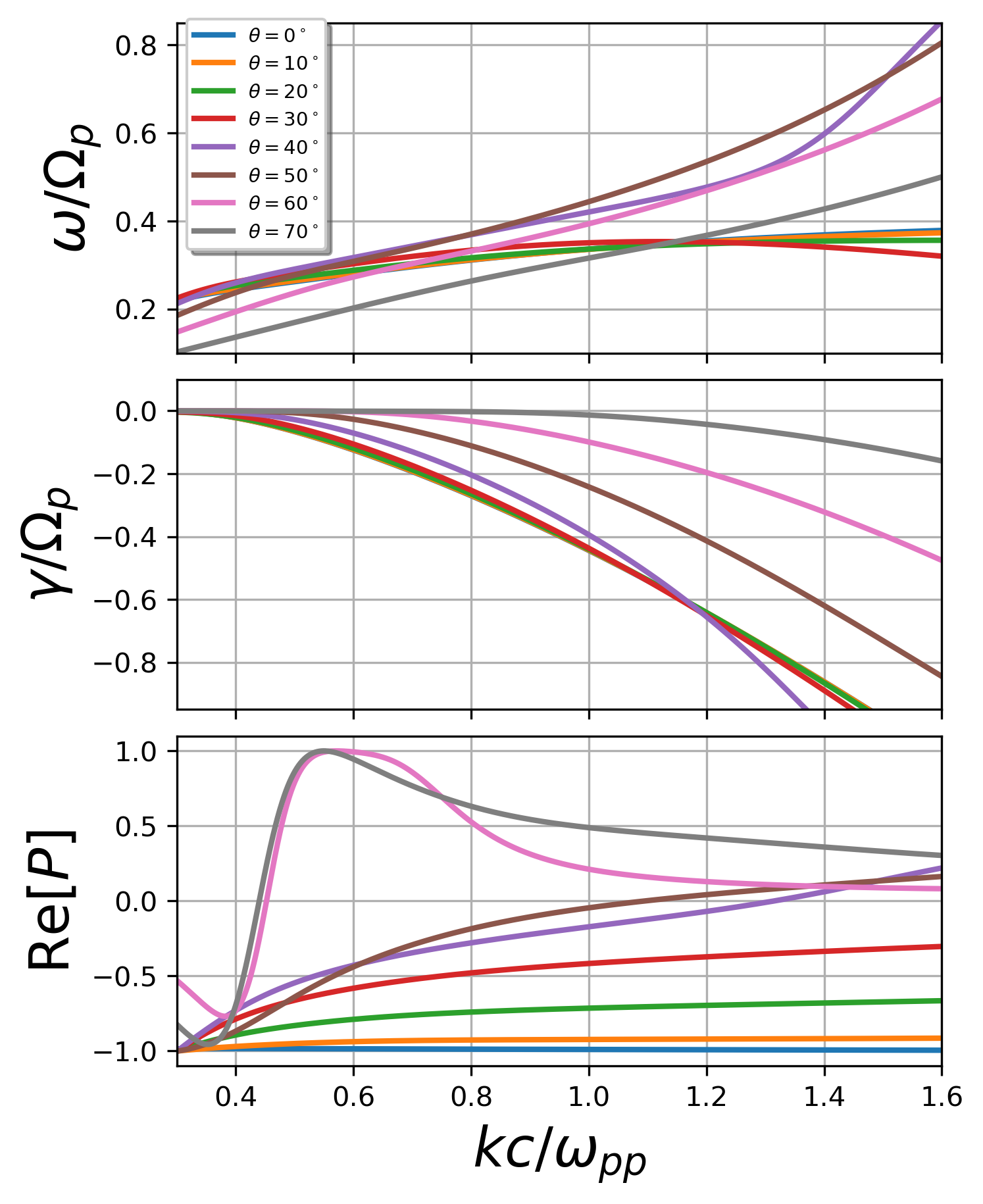}
	\caption{Dispersion relation and polarization spectra for Alfv\'enic waves propagating at several wave-normal angles ($0^\circ \leq \theta \leq 70^\circ$), in an isotropic ($T_\bot = T_\parallel$) electron-proton plasma, with $\beta = 1$, and $C_A = 3.5\,\times 10^{-3}$. From top to bottom panels show real part of the frequency, growth rate and real part of the polarization, respectively, all as function of normalized wave-number. Frequencies are expressed in units of the proton gyrofrequency and each polarization curve is normalized to its maximum value.  
	}
	\label{FIG:1}
\end{figure}

Fig.~\ref{FIG:1} shows the dispersion relation and polarization spectra for Alfv\'enic waves propagating at several wave-normal angles ($0^\circ \leq \theta \leq 70^\circ$). Following the previous considerations about the typical plasma parameters in the inner magnetosphere, here we consider an isotropic ($T_\bot = T_\parallel$) electron-proton plasma, with $\beta = 1$, and $C_A = 3.5\,\times 10^{-3}$. Top panel in the figure shows the real part of the frequency in units of the proton gryrofrequency as function of the wave-number $k = |\mathbf{k}|$, normalized to the proton inertial length. The curve for each wave-normal angle shows that, in general, as the angle increases, the Alfv\'en branch increases its frequency such that at a certain wave-number the solutions changes from the $\omega < \Omega_p$ regime (KAW modes), to $\omega > \Omega_p$ (not shown in the figure) in the so called Kinetic Alfv\'en Whistler as defined by~\citet{Sahraoui2012}, also for an electron-proton plasma with $\beta \sim 1$. Even though this is an interesting behaviour and worth to be analyze elsewhere, here, as we are interested in the change from EMIC to KAW solutions, we only focus to the wave-number region in which $\omega < \Omega_p$. 

On the other hand, regarding the stability of the waves, middle panel in Fig.~\ref{FIG:1} shows the imaginary part of the frequency ($\gamma$) also normalized to the proton gyrofrequency. We can see that, as with this parameter selection (isotropic plasma) there is no free energy to excite Alfv\'enic waves, and therefore $\gamma \leq 0$ for all wave-normal angles. However, it is clear that $\gamma$ increases with increasing angle, which is also consistent with previous studies~\citep[see e.g.][]{Sahraoui2012,Lopez2017}. Further, bottom panel in Fig.~\ref{FIG:1} shows the polarization spectrum for each wave-normal angle. For the sake of identifying the sign of the polarization, the figure shows each curve normalized to its maximum value. Thus, in this fashion the plot allows easily to distinguish KAW modes ($\rm{Re}\,[P]$ $>$ 0) from EMIC waves ($\rm{Re}\,[P]$ $<$ 0). As shown by~\citet{gary1986}, there is a transition wave-normal angle that allows the Alfv\'en branch to shift from EMIC to KAW. In this case, we can see that the Alfv\'enic solutions correspond to KAWs only for $\theta \ge 40^\circ$, and that the wave-number region in which the polarization is positive increases with increasing angle. 

\begin{figure}
	\centering
		\includegraphics[width=.75\linewidth]{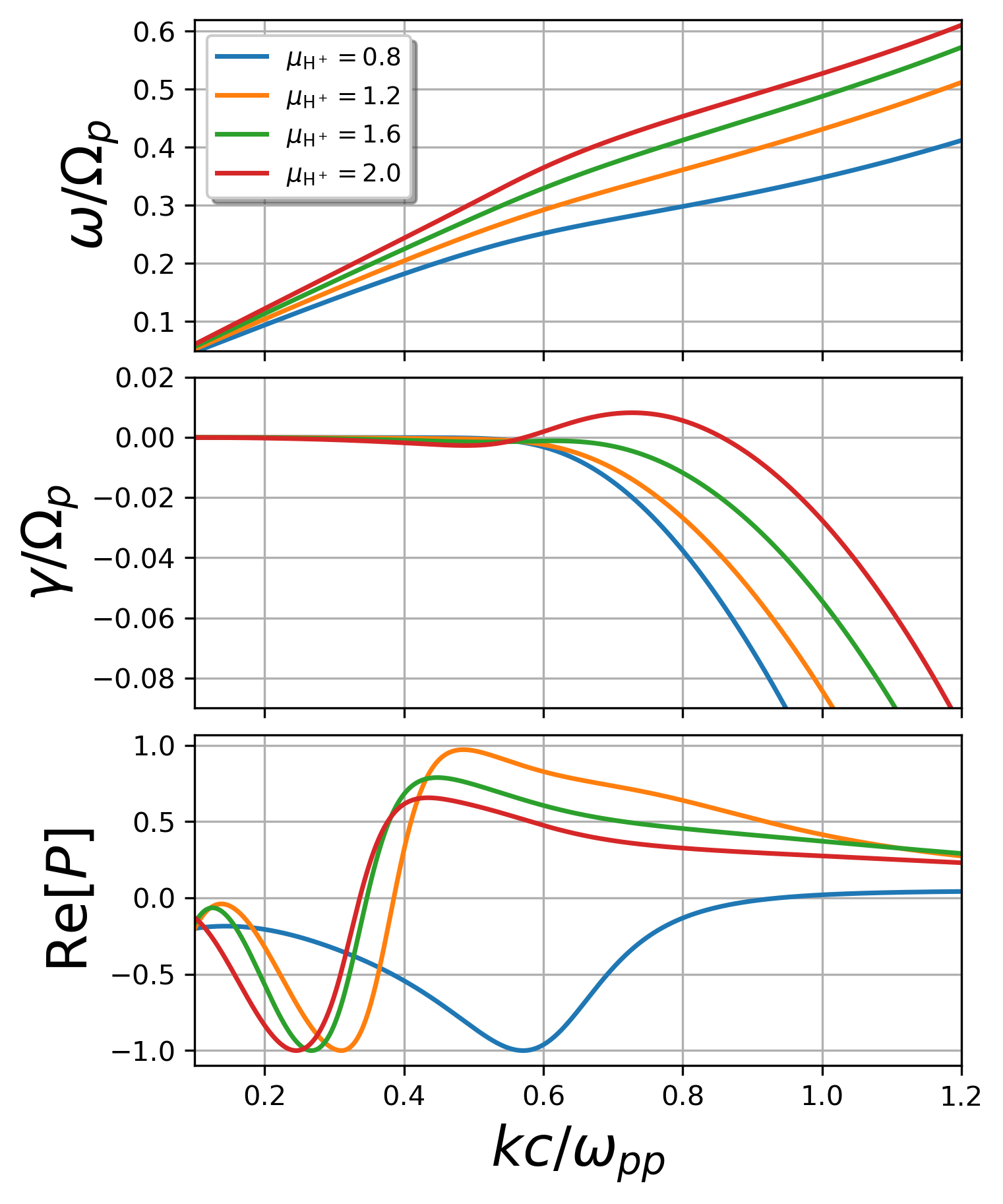}
	\caption{Same as Fig.~\ref{FIG:1} but considering isotropic electrons and protons with different temperature anisotropy values ($0.8 \leq \mu_{\rm{H^+}} \leq 2.0$). In all cases the wave-normal angle is $\theta$ = 60$^\circ$, $\beta_\parallel = 1$ for both species, and $C_A = 3.5\,\times 10^{-3}$. From top to bottom panels show real part of the frequency, growth rate and real part of the polarization, respectively, all as function of normalized wave-number. Frequencies are expressed in units of the proton gyrofrequency, and each polarization curve is normalized to its maximum value.
	}
	\label{FIG:2}
\end{figure}

 As expected, from Fig.~\ref{FIG:1} we observe that the waves behave as KAW modes instead of EMIC waves for larger wave-normal angle and wave-numbers. However, to analyze the stability of the plasma to the propagation of KAWs, we need to consider unstable distributions. This is shown in Fig.~\ref{FIG:2}, in which we present the real part (top panel), and imaginary part (middle panel) of the frequency, and polarization (bottom panel), all as function of wave-number. We use the same plasma parameters as in Fig.~\ref{FIG:1} but now we consider anisotropic protons with different temperature anisotropy values between $\mu_{\rm{H^+}} = 0.8$ and $\mu_{\rm{H^+}} = 2.0$. In all cases the wave-normal angle is $\theta$ = 60$^\circ$, which from Fig.~\ref{FIG:1} we already knew that is a propagation angle that allows the Alfv\'en branch to behave as KAW. From the figure we can see that the dispersion curve rises to larger frequencies as the anisotropy increases (top panel). At the same time, as expected, the imaginary part (middle panel in Fig.~\ref{FIG:2}) also increases with anisotropy, with the mode becoming unstable only for $\mu_{\rm{H^+}} = 2.0$, with a maximum growth rate $\gamma/\Omega_p \sim 0.01$ at $kc/\omega_{pp} \sim 0.75$. Finally, bottom panel in Fig.~\ref{FIG:2} shows the corresponding polarization for each temperature anisotropy value. The figure shows that, even though all $\mu_{\rm{H^+}}$ values allow KAW solutions in a certain $kc/\omega_{pp}$ range, the region in which the mode is right-handed ($\rm{Re}\,[P]$ > 0) depends on the anisotropy of protons, increasing with increasing $\mu_{\rm{H^+}}$. This is expected as at fixed wave-normal angle and fixed $\beta_{\parallel \rm{H^+}} = 1$, the proton gyroradius $ \rho_{\rm{H^+}}$ decreases with decreasing anisotropy, and therefore a larger $k$ (or $k_\perp$) is needed to fulfill the $k_{\perp} \rho_{\rm{H^+}} \sim 1$ condition for KAW modes. In this case from Fig.~\ref{FIG:2} we see that for $\mu_{\rm{H^+}}$ = 0.8 the polarization is positive only for $kc/\omega_{pp}$ > 1.0. In summary, combining the information of the three panels in Fig.~\ref{FIG:2} it is possible to observe that at the wave-numbers at which the mode is unstable ($kc/\omega_{pp} \sim 0.75$) the modes have frequency $\omega < \Omega_p$ and positive polarization. Thus, we conclude that an electron-proton plasma with $\beta \sim 1$ allows the propagation of unstable KAWs.
 
\subsection{The effect of O$^+$ ions}

To analyze the effect of the heavy ions on the existence of unstable KAW, now we introduce O$^+$ ions and repeat the same calculations as in the previous section. As already mentioned, we fix $\bar\beta_{\parallel s} = 1$ for all species and vary the wave-normal angle, and the abundance of the O$^+$ ions. Also, to study the stability of the Alfv\'enic waves, we also vary the temperature anisotropy of protons $\mu_{\rm{H^+}}$ considering values in which the electron-proton plasma is unstable to Alfv\'enic (EMIC or KAW) modes. In addition, for a better comparison with the electron-proton case, from the two Alfv\'enic solutions of the dispersion relation we only analyze the proton band $\omega > \Omega_{\rm{O^+}}$, and avoid the  O$^+$ band due to very restricted frequency limits ($0 < \omega < \Omega_{\rm{O^+}} \approx \Omega_p/16$). 

\begin{figure}
	\centering
		\includegraphics[width=0.75\linewidth]{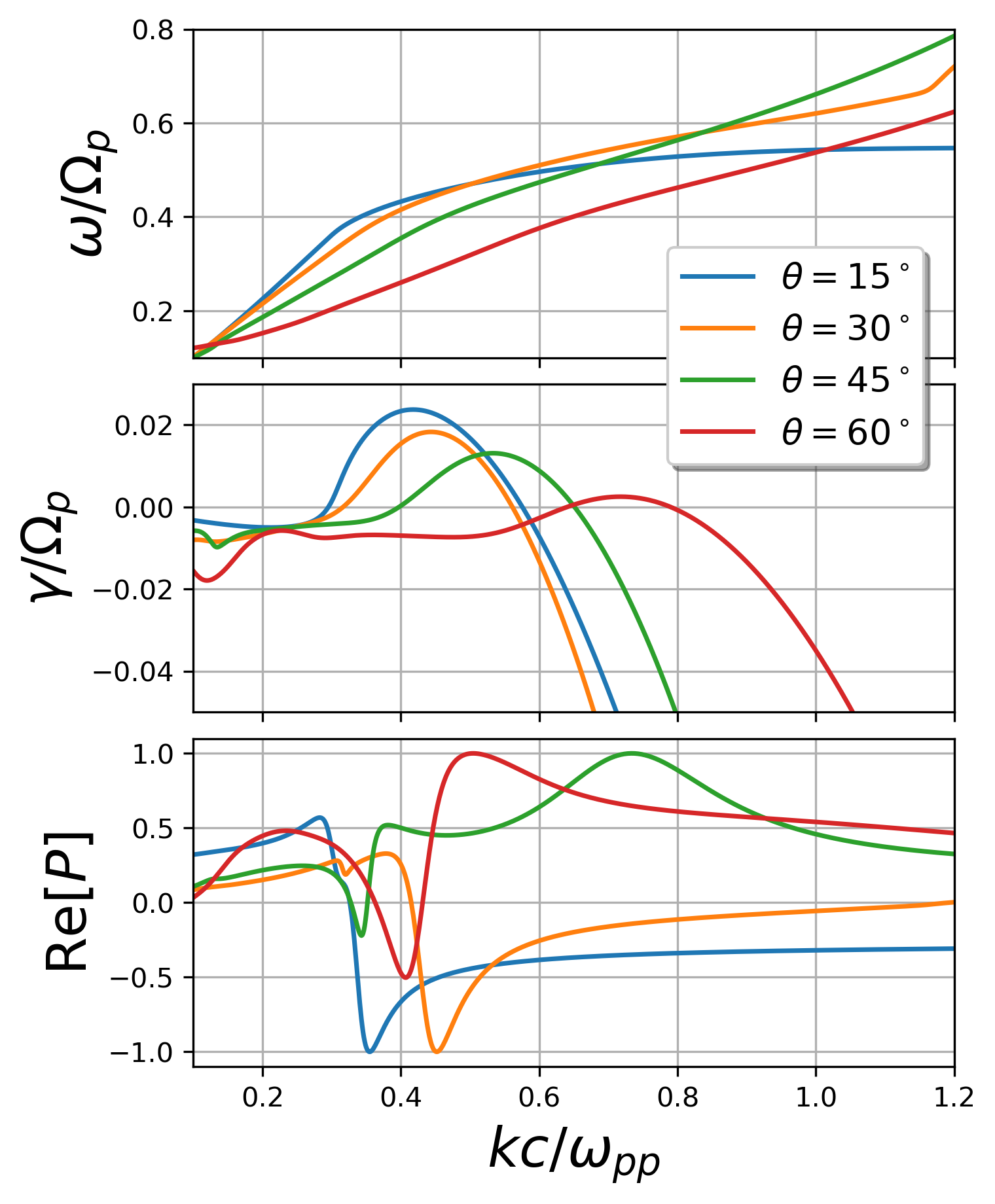}
	\caption{Dispersion relation and polarization spectra for Alfv\'enic waves propagating at four different wave-normal angles ($15^\circ \leq \theta \leq 60^\circ$), in a plasma composed by electrons, 90\% of anisotropic ($\mu_{\rm{H^+}} = 2.0$) H$^+$ ions, and 10\% O$^+$ ions ($\eta_{\rm{O^+}} = 0.1$). For all species $\bar\beta_{\parallel s} = 1$, and $C_A = 3.5\,\times 10^{-3}$. From top to bottom panels show real part of the frequency, growth rate and real part of the polarization, respectively, all as function of normalized wave-number. Frequencies are expressed in units of the proton gyrofrequency and each polarization curve is normalized to its maximum value.
	}
	\label{FIG:3}
\end{figure}

Fig.~\ref{FIG:3} shows the dispersion relation and polarization spectra for Alfv\'enic waves propagating at four different wave-normal angles ($15^\circ \leq \theta \leq 60^\circ$), in a plasma composed by electrons, 90\% of anisotropic ($\mu_{\rm{H^+}} = 2.0$) H$^+$ ions, and 10\% O$^+$ ions ($\eta_{\rm{O^+}} = 0.1$). As in previous calculations, for all species we set $\bar\beta_{\parallel s} = 1$, and $C_A = 3.5\,\times 10^{-3}$. From top to bottom panels show real part of the frequency, growth rate and real part of the polarization, respectively, all as function of normalized wave-number. On the top panel we can see that the presence of O$^+$ ions does not modify much the real part of the frequency. The behavior is similar to the electron-proton case (see Fig.~\ref{FIG:1}) and the mode transits from a more EMIC wave shape at $\theta =15^\circ$, towards the KAW mode ($\omega < \Omega_p$) and Kinetic Alfv\'en Whistler ($\omega > \Omega_p$) at more oblique angles. From Fig.~\ref{FIG:3} we also observe that the maximum growth rate decreases with increasing angle (see middle panel), and that the presence of only a 10\% of isotropic O$^+$ ions implies an important reduction of the maximum growth of the waves. In this case the maximum growth rate at $\theta = 60^\circ$ is $\gamma/\Omega_p \sim 0.005$, which is about half compared to the electron-proton case with the same proton temperature anisotropy ($\mu_{\rm{H^+}} = 2.0$) shown in Fig.~\ref{FIG:2}. On the contrary, regarding polarization spectra, bottom panel in Fig.~\ref{FIG:3} shows that including O$^+$ ions tends to enlarge the wave-number region in which the polarization is positive, allowing the existence of KAW in a wider wave-number range and at smaller wave-normal angles compared to the electron-proton case. In addition, from Fig.~\ref{FIG:3} we also note that for $\theta$ = 45$^\circ$, and 60$^\circ$, the polarization and the growth rate of the waves are positive at the same time. Thus, even though a small abundance of oxygen ions tends to reduce the instability of the waves, a plasma with 10\% of O$^+$ ions also allows the propagation of unstable Kinetic Alfv\'en Waves.   

\begin{figure*}
	\centering
		\includegraphics[width=.95\linewidth]{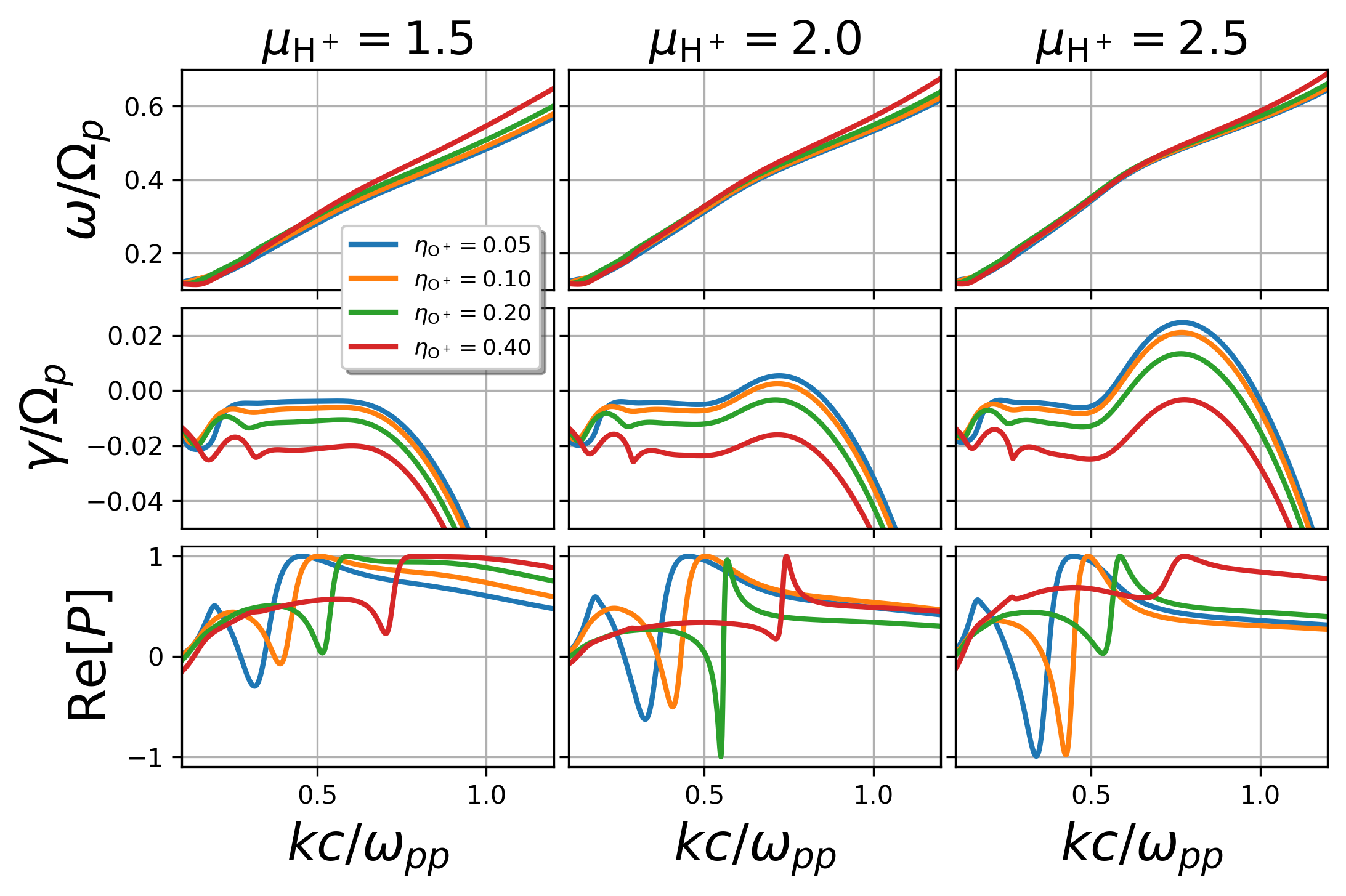}
	\caption{Same as Fig.~\ref{FIG:3} but considering isotropic electrons, isotropic O$^+$ ions with different abundances ($0.05 \leq \eta_{\rm{O^+}} \leq  0.4$), and H$^+$ ions with three different temperature anisotropy values: $\mu_{\rm{H^+}} = 1.5$ (left), $\mu_{\rm{H^+}} = 2.0$ (center), and $\mu_{\rm{H^+}} = 2.5$ (right). In all cases the wave-normal angle is $\theta$ = 60$^\circ$, $\bar\beta_\parallel = 1$ for all species, and $C_A = 3.5\,\times 10^{-3}$. In each column, from top to bottom panels show real part of the frequency, growth rate and real part of the polarization, respectively, all as function of normalized wave-number. Frequencies are expressed in units of the proton gyrofrequency and each polarization curve is normalized to its maximum value.
	}
	\label{FIG:4}
\end{figure*}

A more general view of the role of O$^+$ ions can be seen in Fig.~\ref{FIG:4}, in which we show the same calculations as in Fig.~\ref{FIG:3}, but considering H$^+$ ions with three different temperature anisotropy values: $\mu_{\rm{H^+}} = 1.5$ (left panels), $\mu_{\rm{H^+}} = 2.0$ (center panels), and $\mu_{\rm{H^+}} = 2.5$ (right panels). In addition, for each case we consider 4 different values for the O$^+$ ions abundance ($0.05 \leq \eta_{\rm{O^+}} \leq  0.4$). In all cases the wave-normal angle is $\theta$ = 60$^\circ$, $\bar\beta_\parallel = 1$ for all species, and $C_A = 3.5\,\times 10^{-3}$. The real part of the frequency for all cases is shown in the top panels of Fig.~\ref{FIG:4}. We observe that proton temperature anisotropy and O$^+$ abundance have both mild effect on the real frequency of the Alfv\'en branch (Kinetic Alf\'ven Wave for $\omega < \Omega_p$  and Kinetic Alfv\'en Whistler when $\omega > \Omega_p$). Nevertheless, all dispersion curves show that $\omega$ increases with increasing $\mu_{\rm{H^+}}$ or $\eta_{\rm{O^+}}$. However, this is not the case for the imaginary part of the frequency (middle row in Fig.~\ref{FIG:4}), in which both quantities play opposite roles; i.e., the temperature anisotropy of protons enhances the instability, and the abundance of O$^+$ ions diminishes the growth rates. Moreover, even for the most unstable case ($\mu_{\rm{H^+}}=2.5$) for $\eta_{\rm{O^+}} = 0.4$ the presence of the O+ ions completely inhibits the instability at all wave-numbers. As mentioned in the introduction section, these results were expected as the growth rates of resonant micro-instabilities depend on the relative abundance of the plasma particles undergoing resonant wave-particle interactions. In this case, only resonances between waves and anisotropic protons contribute to the instability, whereas resonant or non-resonant wave-particle interactions with the isotropic O$^+$ population cannot transfer free energy from the plasma to the electromagnetic fields. Finally, bottom panels in Fig.~\ref{FIG:4} show the polarization spectra for each combination between H$^+$ anisotropy and O$^+$ abundance. As in the previous cases shown in Fig.~\ref{FIG:3}, the region in which the mode is right-hand polarized ($\rm{Re}[P] > 0$ increases its extent with increasing $\eta_{\rm{O^+}}$, so that a large enough O$^+$ abundance leads to KAW solutions at almost all considered wave-numbers.

\section{Summary and Conclusions}
\label{ref:conclusions}

In this article, using Vlasov linear theory of plasma waves we have studied the dispersion properties of Alfv\'enic modes in two different plasma environments: first, an electron-proton plasma; and second, a multi-species plasma composed by electrons, protons, and also O$^+$ ions, considering macroscopic plasma parameters relevant to the inner magnetosphere. Using a numerical dispersion solver for linear plasma waves we obtained the dispersion relation considering both cases (electron-proton and multi-species), and analyzed the changes introduced by the O$^+$ ions in the dispersion relation, stability and polarization of Alfv\'enic modes with frequencies similar to the proton gyrofrequency. 

Motivated by in situ measurements of KAW modes, that have been directly observed in the multi-species inner magnetosphere propagating at about $60^\circ$~\citep{chaston2014,moya2015}, we looked for the condition under which the Alfv\'enic modes correspond to unstable right-hand polarized KAW, and then we introduced O$^+$ ions to compare the results of the electron-proton plasma and the multi-species case. To do so we have considered wave-normal angles not too close to $90^\circ$, because in such case the KAW nature of the solutions is granted for almost any value of the plasma $\beta$~\citep{gary1986}. In addition, in such case Bernstein modes will also appear~\citep[see e.g.][]{Lopez2017}. These modes should also play a role on the dynamics of the system, and are worth to study, but such calculations, as well as other effects like temperature anisotropy or drift velocity of the O+ ions, are beyond the scope of our current study, and we will consider them in a subsequent analysis. 

First, regarding the electron-proton case, our numerical results showed that the waves behave as KAW modes instead of EMIC waves for large enough wave-normal and large enough wave-number, and that the unstable modes have frequency $\omega < \Omega_p$ and positive polarization at the same wave-number interval. Thus, through this first analysis we were able to corroborate previous theoretical~\citep{gary1986,hollweg1999,Gary2004} and observational~\citep{chaston2014,moya2015} results, validate our dispersion code, and also provide a framework to compare with the multi-species case. We conclude that an electron-proton plasma with $\beta \sim 1$ allows the propagation of unstable KAWs at $\theta\sim 60^\circ$, as expected. On the other hand, compared to an electron-proton plasma, we have shown that the presence of O$^+$ ions allows the existence of Kinetic Alfv\'en Waves in a larger wave-number range. However, the isotropic O$^+$ ions tend to drastically reduce the growth rates of unstable KAW triggered by anisotropic protons, or even inhibit the instability if their relative abundance is large enough.  

In summary, we have shown that for plasma parameters relevant for the inner magnetosphere, the triggering and propagation Kinetic Alfv\'en Waves due to anisotropic protons is determined by the abundance of O$^+$. Thus, as in the inner magnetosphere O$^+$ ions can even dominate the plasma composition during geomagnetically active times, our results suggest that O$^+$ ions may play an important role on the processes mediated by the wave-particle interactions between plasma particle and Kinetic Alfv\'en Waves, such as the energy transfer from large macroscopic scales to sub-ionic and electron scales. We plan to address these aspects and other non-linear issues, expanding the scope of our model in subsequent works. Furthermore, as collisionless plasmas tend to support macroscopic states in which a given combination of plasma beta and temperature anisotropy allows the plasma to remain near or below marginal stability~\citep[see e.g.][]{gary1993}, these results seem to indicate that heavy ions should play an important role on the stability of a multi-species plasma. In the case of the inner magnetosphere, depending their temperature anisotropy, even a small percentage of O$^+$ ions may completely transform the shape of temperature anisotropy and beta kinetic instability thresholds that are usually computed considering only electrons and protons. 

\subsection*{Acknowledgments}

\noindent We acknowledge Victor Pinto for useful discussion. We are grateful for the support of ANID, Chile through FONDECYT grant No. 1191351 (PSM) and National Doctoral Scholarships Nos. 21182002 (IGM), and 21181965 (BZQ). PSM thanks the support of KU Leuven through the BOF Network Fellowship NF/19/001. PSM also would like to thank Gian Luca Delzanno and the Local Organizing Committee of the 2019 "The Plasma Physics of the Magnetosphere" Conference, held in Bra-Pollenzo, Italy, for a great and fruitful scientific meeting, in which an early version of this work was presented.


\end{document}